\documentclass[aps,preprint]{revtex4}%
\usepackage{amsfonts}
\usepackage{amsmath}
\usepackage{amssymb}
\usepackage{graphicx}%
\providecommand{\U}[1]{\protect\rule{.1in}{.1in}}

\begin{document}
\title{Exploring Bell nonlocality of quantum networks with stabilizing and logical operators}
\author{Li-Yi Hsu}
\affiliation{Department of Physics, Chung Yuan Christian University, Chungli, 32081, Taiwan}
\author{Ching-Hsu Chen}
\affiliation{Department of Electrophysics, National Chiayi University, Chiayi, 600, Taiwan}

\begin{abstract}
In practical quantum networks, a variety of multiqubit stabilized states
emitted from independent sources are distributed among the agents, and the
correlations across the entire network can be derived from each agent's local
measurements on the shared composite quantum systems. To reveal the Bell
nonlocality in such cases as a quantum feature, minimal knowledge of the
emitted stabilizer state is required. Here, we demonstrate that knowing the
stabilizing and logical operators indeed provides a new way of exploring Bell
nonlocality in quantum networks. For the qubit distribution in quantum
networks, the associated nonlinear Bell inequalities are derived. On the other
hand, to violate these inequalities, one can design local incompatible
observables\ using minimal knowledge of the emitted states. The tilted
nonlinear Bell inequalities tailored for specific nonmaximal entangled
stabilizer states and a way of achieving the maximal violation are also explored.

\end{abstract}
\volumeyear{year}
\volumenumber{number}
\issuenumber{number}
\eid{identifier}
\date[Date text]{date}
\received[Received text]{date}

\revised[Revised text]{date}

\accepted[Accepted text]{date}

\published[Published text]{date}

\maketitle

\section{Introduction}

In the seminal work of John Bell, it was shown that the quantum correlations
arising from spatially separated systems can break the limits of classical
causal relations \cite{Bell,EPR}. In classical physics, the realism and
locality of spacelike events constrain the strength of classical correlations
bounded by the Bell inequalities. Quantum theory inconsistent with local
realism predicts stronger correlations that violate the Bell inequalities.
Thanks to quantum information science, two-particle and multi-particle quantum
correlations have been extensively investigated. As a distinct feature from
classical physics, Bell nonlocality has led to applications in quantum
information processing including private random number generation
\cite{random}, quantum cryptography \cite{DIC}, and reductions in
communication complexity \cite{complexity}.

In typical Bell experiments on statistical correlations, a source emits a
state comprising two or more particles that are shared between two or more
distant observers, who each perform local measurements with a random chosen
setting and then obtain the measurement outcomes. To reveal the Bell
nonlocality of an entangled state, the local observables should be set
deliberately. For example, to violate the Bell inequalities tailored to
stabilizer states such as graph states, it is helpful to take the stabilizers
as a reference for finding suitable measurement settings \cite{gf1,gf2,gf3}.
To emphasize the role of stabilizing operators and logical bit-flip operators,
we review the Clauser-Horne-Shimony-Holt (CHSH) inequality and its
modification below. Let the two-qubit entangled state ($0<\phi<\frac{\pi}{2}%
$)
\begin{equation}
\left\vert \phi\right\rangle =\cos\phi\left\vert \overline{0}\right\rangle
+\sin\phi\left\vert \overline{1}\right\rangle \label{psi}%
\end{equation}
be the codeword of the [[2, 1, 2]] stabilizer-based quantum error-detecting
code with the logical states $\left\vert \overline{0}\right\rangle =\left\vert
00\right\rangle $ and $\left\vert \overline{1}\right\rangle =\left\vert
11\right\rangle $. The stabilizer generator, logical bit-flip, and phase-flip
operators are $\sigma_{z}\otimes$ $\sigma_{z}$, $\sigma_{x}\otimes$
$\sigma_{x}$, and $I\otimes\sigma_{z}$, respectively ($\sigma_{x}$,
$\sigma_{y}$, and $\sigma_{z}$ are Pauli matrices and $I$ is the identity
operator). In the bipartite Bell test, the CHSH operator is $\mathbf{B}%
_{CHSH}=\sum\nolimits_{i,\text{ }j=0}^{1}(-1)^{ij}A_{i}\otimes B_{j}$, where
$A_{i}$ and $B_{j}$ are local observables, and the CHSH inequality states that
$\left\langle \mathbf{B}_{CHSH}\right\rangle \overset{c}{\leq}2$, where the
$\left\langle \cdot\right\rangle $ denotes the expectation value of $\cdot.$
To violate the CHSH inequality, for the first-qubit, we assign
\begin{equation}
\sigma_{z}\rightarrow\frac{1}{2\cos\mu}(A_{0}+A_{1})\text{, }\sigma
_{x}\rightarrow\frac{1}{2\sin\mu}(A_{0}-A_{1}),\label{a}%
\end{equation}
and for the second-qubit, we assign
\begin{equation}
\sigma_{z}\rightarrow B_{0}\text{, }\sigma_{x}\rightarrow B_{1\text{ }%
}\label{b}%
\end{equation}
It is easy to verify that
\begin{align}
&  \left\langle \phi\left\vert \mathbf{B}_{CHSH}\right\vert \phi\right\rangle
\nonumber\\
&  =2\cos\mu\left\langle \phi\left\vert \sigma_{z}\otimes\sigma_{z}\right\vert
\phi\right\rangle +2\sin\mu\left\langle \phi\left\vert \sigma_{x}\otimes
\sigma_{x}\right\vert \phi\right\rangle \nonumber\\
&  =2(\cos\mu+\sin\mu\sin2\phi)\nonumber\\
&  =2\sqrt{1+\sin^{2}2\phi}\cos(\mu-\mu_{0})\nonumber\\
&  \leq2\sqrt{1+\sin^{2}2\phi},\label{CHSH}%
\end{align}
where $\tan\mu_{0}=\sin2\phi$. Some remarks are in order. First, the operators
$(A_{0}+A_{1})\otimes B_{0}=\cos\mu\sigma_{z}\otimes\sigma_{z}$ and
$(A_{0}-A_{1})\otimes B_{1}=\sin\mu\sigma_{x}\otimes\sigma_{x}$, and hence
both terms in the first equality in (\ref{CHSH}) exemplify the usefulness of
the stabilizing operator and the logical (bit-flip) operator for finding the
local observables that violate the CHSH inequality. The maximal CHSH value
larger than two can be achieved by setting $\mu=\mu_{0}$. In particular,
$\left\vert \phi=\frac{\pi}{4}\right\rangle $ is maximally entangled,
$\sigma_{x}\otimes$ $\sigma_{x}$ becomes another stabilizing operator rather
than simply a logical bit-flip operator, and the CHSH inequality can be
maximally violated. There are a variety of Bell inequalities violated by the
graph states as a specific family of stabilizer states, where the associated
Bell operators can be reformulated as the sum of their stabilizing operators,
and hence the perfect/antiperfect correlations therein can reach the maximal
violation \cite{gf2,gf3,gs2,gs3,gs4,gs5}. If the multi-qubits mixed states
involve two stabilizing operators, their nonlocality can be verified by the
violation of stabilizer-based Bell-type inequalities \cite{mix1,mix2}. Second,
observables $A_{0}$ and $A_{1}$ can be regarded as the results of
\textquotedblleft cutting and mixing\textquotedblright\ the stabilizing
operator and the logical bit-flip operator into the local observables:
\begin{equation}
\sigma_{z}\otimes\sigma_{z},\sigma_{x}\otimes\sigma_{x}\overset{\text{cutting}%
}{\rightarrow}\sigma_{z},\sigma_{x}\overset{\text{mixing}}{\rightarrow
}A_{x_{i}}=\cos\mu\sigma_{z}+(-1)^{_{x_{i}}}\sin\mu\sigma_{x}.\label{mix}%
\end{equation}
Cutting means cutting the first qubit observables from $\sigma_{z}%
\otimes\sigma_{z}$ and $\sigma_{x}\otimes\sigma_{x}$ ; mixing means linearly
superposing the two cut observables. In what follows, the local observables on
the source side will be constructed in a similar way. In addition, note that
the local observables anticommute; i.e., $\left\{  A_{0}\text{, }%
A_{1}\right\}  =\left\{  B_{0}\text{, }B_{1}\right\}  =0$. Third, the
observable $B_{0}$ is the phase-flip operator, which can be exploited in the
tilted CHSH operators $\mathbf{B}_{\beta\text{-}CHSH}=\beta B_{0}+$
$\mathbf{B}_{CHSH}$ with the Bell inequality $\left\langle \mathbf{B}%
_{\beta\text{-}CHSH}\right\rangle \overset{c}{\leq}2+\beta$. By setting
$\beta=2/\sqrt{1+2\tan^{2}2\phi}$, the maximal violation can be achieved\ by
$\left\vert \phi\right\rangle $ \cite{non1, non2}. That is, the nonmaximally
entangled state can maximally violate the tilted Bell inequalities involving
the phase-flip operators.

One can generalize the above approach to the multi-qubit case. For example,
let $\left\vert \psi\right\rangle =\cos\psi\left\vert \overline{0}%
\right\rangle +\sin\psi\left\vert \overline{1}\right\rangle $ be the codeword
of the [[$5$, $1$, $3$]] stabilizer-based quantum error-correcting code
(SQECC), and let Alice hold the first qubit and Bob hold the other qubits. In
this case, the useful stabilizing operator and logical bit-flip operator are
$g=\sigma_{z}\otimes\sigma_{z}\otimes\sigma_{x}\otimes I\otimes\sigma_{x}$ and
$\sigma_{x}^{\otimes5}$, respectively. The observables of the first and second
qubits of the stabilizing operators and the bit-flip operator are the same as
those in (\ref{a}) and (\ref{b}), respectively. The last three qubits are
termed\textit{ idle} qubits since the observable $\sigma_{x}$ is always
measured on them, while the outcome of the fourth qubit is discarded if Bob
measures $B_{1}^{\prime}$. In the Bell test, Alice randomly measures one of
the observables in (\ref{a}), whereas Bob randomly measures $B_{0}^{\prime
}=B_{0}\otimes\sigma_{x}^{\otimes3}$ or $B_{1}^{\prime}=B_{1}\otimes\sigma
_{x}\otimes I\otimes\sigma_{x}$. The CHSH-like inequality can be written as
$\left\langle \sum\nolimits_{i,\text{ }j=0}^{1}(-1)^{ij}A_{i}\otimes
B_{j}^{\prime}\right\rangle \overset{c}{\leq}2.$ As another example, let the
logical states be $\left\vert \overline{0}\right\rangle =\left\vert
0\right\rangle ^{\otimes n}$ and $\left\vert \overline{1}\right\rangle
=\left\vert 1\right\rangle ^{\otimes n}$, and denote the $n$-qubit
Greenberger--Horne--Zeilinger (GHZ) state ($n\geq3$) as $\left\vert GHZ_{\phi
}\right\rangle =\cos\phi\left\vert \overline{0}\right\rangle +\sin
\phi\left\vert \overline{1}\right\rangle $. Alice holds the first $m$ qubits
and Bob holds the other $n-m$ qubits. Here, the useful stabilizing, logical
bit-flip and phase-flip operators can be set as $\sigma_{z}\otimes I^{\otimes
m-1}\otimes\sigma_{z}\otimes I^{\otimes n-m-1}$, $\sigma_{x}^{\otimes n}$, and
$I^{\otimes m}\otimes\sigma_{z}\otimes I^{\otimes n-m-1}$, respectively. One
can construct the local observables $A_{i}^{\prime\prime}=\cos\mu\sigma
_{z}\otimes I^{\otimes m-1}+(-1)^{i}\sin\mu\sigma_{x}^{\otimes m}$ using the
cutting-and-mixing scheme, and,\ $B_{j}^{\prime\prime}=(1-j)\sigma_{z}\otimes
I^{\otimes n-m-1}+j\sigma_{x}^{\otimes n-m}$ such that $\left\{  A_{0}%
^{\prime\prime}\text{, }A_{1}^{\prime\prime}\right\}  =\left\{  B_{0}%
^{\prime\prime}\text{, }B_{1}^{\prime\prime}\right\}  =0$. Hence, we reach the
CHSH-like inequality $\left\langle \sum\nolimits_{i,\text{ }j=0}^{1}%
(-1)^{ij}A_{i}^{\prime\prime}\otimes B_{j}^{\prime\prime}\right\rangle
\overset{c}{\leq}2$. Here, the last $n-m-1$ qubits are idle since only the
observable $\sigma_{x}$ is always measured on each of them. Conditional on the
qubit assignment, one can select the useful stabilizing operators and logical
operators to derive similar CHSH-like inequalities.

Recently, Bell nonlocality in quantum networks as the generalized Bell
experiments have attracted much research attention. Long-distance quantum
networks of large-scale multi-users are the substantial goals of quantum
communication, so it is fundamental to study their nonlocal capacities. A
quantum network involves multiple independent quantum sources, where each of
them initially emits a two- or multi-qubit entangled state shared among a set
of observers or agents. There are several obstacles to the study of
nonlocality in a quantum network. The classical correlations of a network
indicate more sophisticated causal relations and lead to stronger constraints
than those in the typical (one-source) Bell scenario. In addition, each
observer can perform a joint measurement on the qubits at hand, which could
result in strong correlations across the network. Unlike the typical linear
Bell-type inequalities, most Bell-type inequalities for various classes of
networks are nonlinear, which implies the nonconvexity of the multipartite
correlation space \cite{star, noncyclic,Poly,luo1}. In the two-source case as
the simplest quantum network, bilocal and nonlocal correlations were
thoroughly investigated \cite{bilocal1, bilocal2}. Next the Bell-type
inequalities for star-shaped and noncyclic networks were studied \cite{star,
noncyclic}. Recently, broader classes of quantum networks based on locally
causal structures have also been investigated \cite{Poly, Poly1, luo1,luo2}.
In particular, computationally efficient algorithms for constructing Bell
inequalities have been proposed \cite{luo1}. These Bell-type inequalities are
tailored for networks with quantum sources emitting either two-qubit Bell
states or generalized GHZ states. The stabilizing operators and logical
operators implicitly play substantial roles in setting up local joint
observables. On the other hand, regarding the potential applications of
encrypted communication in quantum networks, stabilizer quantum
error-correcting codes (SQECCs) can be more useful in quantum network
cryptography, such as in quantum secret sharing \cite{QSS} and secure quantum
key distribution protocols \cite{Preskill}. Revealing Bell nonlocality in a
network is required for detecting potential eavesdropping attacks. Moreover, a
structured quantum state with stabilizing operators and logical operators is
more useful in the engineering of quantum networks \cite{percolation,
memory,repeater1, QSS1, QSS2}.

In this work, we study the\ Bell nonlocality of quantum networks. Hereafter,
we consider a $K$-locality network $\mathcal{N}$ as shown in Fig. (1). There
are $K+M$ agents of which $K$, $\mathcal{S}^{(1)}$\ldots$\mathcal{S}^{(K)}$,
are on the source side and $M$, $\mathcal{R}^{(1)}$\ldots$\mathcal{R}^{(M)}$,
on the receiver side. There are $N$ independent sources. Let $0=e_{0}%
<e_{1}<\cdots<e_{K}=N$. The agent $\mathcal{S}^{(s)}$ holds the sources
$e_{s-1}+1$,\ldots\ , and $e_{s}$; thus the number of sources that
$\mathcal{S}^{(s)}$ holds is $(e_{s}-e_{s-1})$. The source $i$ ($e_{s-1}+1\leq
i\leq e_{s}$) held by $\mathcal{S}^{(s)}$ emits $n_{i}$ particles of which
$n_{i}^{(0)}$ ($\neq0$) are in possession of the agent $\mathcal{S}^{(s)\text{
}}$and $n_{i}^{(m)}$($\geq$ $0$) are sent to the agent $\mathcal{R}^{(m)}$.
Consequently, $\sum_{j=0}^{M}n_{i}^{(j)}=n_{i}$. In the classical networks,
source $i$ emits $n_{i}$ particles described by hidden state $\lambda_{i}$; in
the quantum networks, it emits $n_{i}$ qubits in quantum state $\left\vert
\psi_{(i)}\right\rangle $, which can be either a stabilizer state or a
codeword of a SQECC. In the Bell test, observer $\mathcal{S}^{(s)}%
$($\mathcal{R}^{(m)}$) measures observable $A_{x_{s}}^{(s)}$ ($B_{y_{m}}%
^{(m)}$) with the associated outcome $a_{s}$ ($b_{m}$), where $x_{s}$,
$y_{m}\in\{0,1\}$ and $a_{s}$, $b_{m}\in\{-1,1\}$. In the following, the index
pair $(i,$ $j)$ denotes the $j$-th particle emitted from source\ $i$, and
$(i$, $j)\rightarrow\mathcal{X}^{(k)}$ indicates that particle $(i$, $j)$ is
at agent $\mathcal{X}^{(k)}$'s hand ($\mathcal{X}\in\{\mathcal{R}$,
$\mathcal{S}\}$). Finally, we denote the particle sets $\mathbb{S}%
^{(k)}=\{(i,j)|\forall(i,j)\rightarrow\mathcal{S}^{(k)}\}$ and $\mathbb{R}%
^{(k)}=\{(i,j)|\forall(i,j)\rightarrow\mathcal{R}^{(k)}\}$.

The remainder of this paper is organized as follows: In Sec. II, we
investigate the classical networks, where general local causal models (GLCMs)
are introduced. Then, Bell inequalities associated with classical networks are
proposed. We study Bell nonlocality of $K$-locality quantum networks in Sec.
III. First we review the stabilizer states and SQECCs. Then we demonstrate how
to violate the proposed Bell inequalities using deliberated local observables.
It will be shown that, the local observables can be made up of
\textquotedblleft cut-graft-mixing\textquotedblright\ stabilizing operators
and logical operators. Notably, there are two alternative nonlocal
correlations. One is due to the entanglement of the logical states themselves,
where two suitable stabilizing operators are exploited to construct the local
observables. The other nonlocality results from the entanglement due to the
superposition of logical states, in which a stabilizing operator and a logical
operator can be suitably exploited in this case. We illustrate these two kinds
of nonlocality in terms of the 5-qubit code, which is the smallest stabilizer
code that protects against single-qubit errors. In Sec. IV, we propose Bell
inequalities tailored for nonmaximally entangled states distributed in a
quantum network. Finally, conclusions are drawn in Sec. V.

\section{Classical networks}

\subsection{General local causal models}

The general local causal models (GLCMs) in classical networks can be described
as follows: The $i$-th source is associated with a local random variable as
the hidden state $\lambda_{i}$ in the measure space $(\Omega_{i}$, $\Sigma
_{i}$, $\mu_{i})$\ with the normalization condition $\int_{\Omega_{i}}d\mu
_{i}(\lambda_{i})\ =1$. All systems in the Bell test scenario are considered
in the hidden state $\Lambda=(\lambda_{i},\cdots,$ $\lambda_{N})$ in the
measure space $(\Omega$, $\Sigma$, $\mu)$, where $\Omega=$ $\Omega_{1}%
\otimes\cdots\otimes\Omega_{N}$ and the measure of $\Lambda$ is given by
$\mu(\Lambda)=\prod\mu_{i}(\lambda_{i})$ with the normalization condition
$\int_{\Omega}d\mu(\Lambda)=1$. In the measurement phase, agent $\mathcal{S}%
^{(s)}$ performs the measurement $A_{x_{s}}^{(s)}$ on state $\Lambda_{S}%
^{(s)}$ with the corresponding outcome denoted by $a_{s}\in\{1,-1\}$.\ The
GLCM suggests a joint conditional probability distribution of the measurement outcomes%

\begin{equation}
P(\mathbf{a}|\mathbf{x})=\int_{\Omega}d\mu(\Lambda)\prod_{s=1}^{K}%
P(a_{s}|x_{s},\Lambda), \label{JP}%
\end{equation}
where $\mathbf{a}=(a_{1},a_{2}$, $\cdots$, $a_{K})$ and\ $\mathbf{x}%
=(x_{1},x_{2}$, $\cdots$, $x_{K})$; hence, we have the correlation%

\begin{equation}
\left\langle A_{x_{1}}^{(1)}\ldots A_{x_{K}}^{(K)}\right\rangle =\sum
\nolimits_{\mathbf{a}}a_{1}\cdots a_{K}P(\mathbf{a}|\mathbf{x},\Lambda).
\end{equation}

On the other hand, in the $K$-locality condition \cite{luo1}, $\mathcal{S}%
^{(s)}$ can access the hidden state $\Lambda_{S}^{(s)}=(\lambda_{e_{s-1}+1}%
$,$\cdots$, $\lambda_{e_{s}})$ in the measure space $\Omega_{S}^{(s)}%
=\Omega_{e_{s-1}+1}\otimes\cdots\otimes\Omega_{e_{s}}$, where
\begin{equation}
\Lambda_{S}^{(s)}\cap\Lambda_{S}^{(s^{\prime})}=\emptyset\text{ }%
\Leftrightarrow\text{ }s\neq s^{\prime}%
\end{equation}
and
\begin{equation}
\cup_{i=1}^{K}\Lambda_{S}^{(i)}=\Lambda.
\end{equation}
Eq. (\ref{JP}) can be rewritten as
\begin{equation}
P(\mathbf{a}|\mathbf{x})=\prod_{s=1}^{K}\int_{\Omega_{S}^{(s)}}d\mu
_{s}(\Lambda_{S}^{(s)})P(a_{s}|x_{s}\text{, }\Lambda_{S}^{(s)}),
\end{equation}
Denote the local expectation as
\begin{equation}
\left\langle A_{x_{s}}^{(s)}\right\rangle =\sum_{a_{s}=-1,1}P(a_{s}%
|x_{s}\text{, }\Lambda_{S}^{(s)}),
\end{equation}
By $K$-locality with the GLCM, we have
\begin{equation}
\left\langle A_{x_{1}}^{(1)}\ldots A_{x_{K}}^{(K)}\right\rangle =\prod
_{s=1}^{K}\left\langle A_{x_{s}}^{(s)}\right\rangle .
\end{equation}
Denote $\Delta^{\pm}A^{(i)}=\frac{1}{2}(\left\langle A_{0}^{(i)}\right\rangle
\pm\left\langle A_{1}^{(i)}\right\rangle )$. Since $-1\leq\left\langle
A_{x_{i}}^{(i)}\right\rangle \leq1$, we have
\begin{equation}
-1\leq\Delta^{-}A^{(i)},\Delta^{+}A^{(i)}\leq1
\end{equation}
and
\begin{equation}
\left\vert \Delta^{+}A^{(i)}\right\vert +\left\vert \Delta^{-}A^{(i)}%
\right\vert =\max\{\left\vert \left\langle A_{0}^{(i)}\right\rangle
\right\vert ,\left\vert \left\langle A_{1}^{(i)}\right\rangle \right\vert
\}\leq1. \label{delta}%
\end{equation}

On the receiving side, let $n_{_{j}}^{(m)}>0$ if $j\in\{j_{1},\cdots
,j_{_{k_{m}}}\}$ and $n_{_{j}}^{(m)}=0$ otherwise. In this case,
$\mathcal{R}^{(j)}$ receives the hidden states $\Lambda_{R}^{(m)}%
=(\lambda_{j_{1}}$,$\cdots$, $\lambda_{j_{k_{m}}})$ in the measure space
$\Omega_{R}^{(m)}=$ $\Omega_{j_{1}}\otimes\cdots\otimes\Omega_{j_{k_{m}}}$,
where $1\leq j_{1}<j_{2}...<j_{k_{m}}\leq N$. In the measurement phase,
$\mathcal{R}^{(m)}$ performs the measurement $B_{y_{m}}^{(m)},$ $y_{m}%
\in\{0,1\}$, on the state $\Lambda_{R}^{(m)}$ with the corresponding outcome
denoted by $b_{m}\in\{1,-1\}$. We have%

\begin{align}
\left\vert \left\langle B_{y_{m}}^{(m)}\right\rangle \right\vert  &
=\left\vert \int_{\Omega_{R}^{(m)}}\prod_{k}d\mu_{k}(\lambda_{k})\sum
_{b_{m}=-1,1}b_{m}P(b_{m}|y_{m}\text{, }\Lambda_{R}^{(m)})\right\vert
\nonumber\\
&  \leq1. \label{bb}%
\end{align}

\subsection{Bell inequalities}

The correlation strength in the proposed $K$-locality network is evaluated by
two quantities:
\begin{equation}
\mathbf{I}_{K,M}=\frac{1}{2^{K}}\left\langle \prod_{i=1}^{K}\prod_{j=1}%
^{M}(A_{0}^{(i)}+A_{1}^{(i)})B_{0}^{(j)}\right\rangle
\end{equation}
and
\begin{equation}
\mathbf{J}_{K,M}=\frac{1}{2^{K}}\left\langle \prod_{i=1}^{K}\prod_{j=1}%
^{M}(A_{0}^{(i)}-A_{1}^{(i)})B_{1}^{(j)}\right\rangle .
\end{equation}
In the classical scenario, we have%

\begin{align}
&  \left\vert \mathbf{I}_{K,M}\right\vert _{GLCM}\nonumber\\
&  =\frac{1}{2^{K}}\int_{\Omega}\prod_{i=1}^{K}\left\vert \left\langle
A_{0}^{(i)}+A_{1}^{(i)}\right\rangle \right\vert \prod_{j=1}^{M}\left\vert
\left\langle B_{0}^{(j)}\right\rangle \right\vert \prod_{k=1}^{N}d\mu
_{k}(\lambda_{k})\nonumber\\
&  \leq\int_{\Omega}\prod_{i=1}^{K}\left\vert \Delta^{+}A^{(i)}\right\vert
\prod_{k=1}^{N}d\mu_{k}(\lambda_{k})
\end{align}
and
\begin{align}
&  \left\vert \mathbf{J}_{K,M}\right\vert _{GLCM}\nonumber\\
&  =\frac{1}{2^{K}}\int_{\Omega}\prod_{i=1}^{K}\left\vert \left\langle
A_{0}^{(i)}-A_{1}^{(i)}\right\rangle \right\vert \prod_{j=1}^{M}\left\vert
\left\langle B_{1}^{(j)}\right\rangle \right\vert \prod_{k=1}^{N}d\mu
_{k}(\lambda_{k})\nonumber\\
&  \leq\int_{\Omega}\prod_{i=1}^{K}\left\vert \Delta^{-}A^{(i)}\right\vert
\prod_{k=1}^{N}d\mu_{k}(\lambda_{k}),
\end{align}
where the inequalities are from (\ref{bb}). Before proceeding further, two
useful lemmas are introduced as follows:

\textit{Lemma 1 }(Mahler inequality) Let $\alpha_{k}$ and $\beta_{k}$ be
nonnegative real numbers, and let $p$ $\in\mathbb{N}$; then,%

\begin{equation}
\prod\nolimits_{k=1}^{p}\alpha_{k}^{1/p}+\prod\nolimits_{k=1}^{p}\beta
_{k}^{1/p}\leq\prod\nolimits_{k=1}^{p}(\alpha_{k}+\beta_{k})^{1/p}.
\end{equation}
The proof can be found in \cite{star}.

We obtain the following nonlinear Bell inequality:
\begin{align}
&  \left\vert \mathbf{I}_{K,M}\right\vert _{GLCM}^{\frac{1}{K}}+\left\vert
\mathbf{J}_{K,M}\right\vert _{GLCM}^{\frac{1}{K}}\nonumber\\
&  \leq\{\int_{\Omega}\prod_{i=1}^{K}(\left\vert \Delta^{+}A^{(i)}\right\vert
+\left\vert \Delta^{-}A^{(i)}\right\vert )\prod_{k=1}^{N}d\mu_{k}(\lambda
_{k})\}^{\frac{1}{K}}\nonumber\\
&  =\{\int_{\Omega}\prod_{i=1}^{K}(\max\{\left\vert \left\langle A_{x_{i}%
=0}\right\rangle \right\vert ,\text{ }\left\vert \left\langle A_{x_{i}%
=1}\right\rangle \right\vert \})\prod_{k=1}^{N}d\mu_{k}(\lambda_{k}%
)\}^{\frac{1}{K}}\nonumber\\
&  \leq(\int_{\Omega}\prod_{j=k}^{N}d\mu_{k}(\lambda_{k}))^{\frac{1}{K}}=1,
\label{NonBell}%
\end{align}
where the first inequality is from Lemma 1, and the fourth line is a
consequence of (\ref{delta}).

\section{Bell nonlocality of a quantum network}

\subsection{Review of stabilizer states and stabilizer-based quantum
error-correcting code}

Let the state $\left\vert \psi_{(i)}\right\rangle $ emitted from the quantum
source $i$ be an $n_{i}$-qubit stabilizer state. By definition, an $n_{i}%
$-qubit stabilizer state is one that is stabilized by a stabilizer which is a
nontrivial subgroup of the Pauli group. In particular, if $\left\vert
\psi_{(i)}\right\rangle $ as a codeword of [[$n_{i}$, $k_{i}$, $d_{i}$]]
SQECC, denote the last ($k_{i}$-th) logical qubit as $\left\vert \overline
{0}_{i}\right\rangle $ and$\ \left\vert \overline{1}_{i}\right\rangle $ and
the corresponding logical bit- and phase-flip operators as $\overline{X}%
_{(i)}^{\text{ }}$ and $\overline{Z}_{(i)}^{\text{ }}$. Without loss of
generality, we have
\begin{equation}
\left\vert \psi_{(i)}\right\rangle =\sum_{z\in\{0,1\}^{\otimes k_{i}}}%
a_{z}\left\vert \overline{z}\right\rangle =\cos\phi_{i}\left\vert \varphi
_{i}^{0}\right\rangle \left\vert \overline{0}_{i}\right\rangle +\sin\phi
_{i}\left\vert \varphi_{i}^{1}\right\rangle \left\vert \overline{1}%
_{i}\right\rangle , \label{SQECCC state}%
\end{equation}
where $\left\langle \varphi_{i}^{0}|\varphi_{i}^{1}\right\rangle \in R$, and
$\left\langle \varphi_{i}^{0}|\varphi_{i}^{0}\right\rangle =\left\langle
\varphi_{i}^{1}|\varphi_{i}^{1}\right\rangle =1$. In addition,%

\begin{equation}
\overline{X}_{(i)}^{\text{ }}\left\vert \overline{0}_{i}\right\rangle
=\left\vert \overline{1}_{i}\right\rangle ,\overline{X}_{(i)}^{\text{ }%
}\left\vert \overline{1}_{i}\right\rangle =\left\vert \overline{0}%
_{i}\right\rangle ,
\end{equation}%
\begin{equation}
\overline{Z}_{(i)}^{\text{ }}\left\vert \overline{c}_{i}\right\rangle
=(-1)^{c}\left\vert \overline{c}_{i}\right\rangle (c\in\{0,1\}).
\end{equation}
In what follows, we exploit $g_{(i)}$ and $g_{(i)}^{\prime}$ as useful
stabilizing operators, and $\overline{X}_{(i)}^{\text{ }}$ and $\overline
{Z}_{(i)}^{\text{ }}$ as useful logical operators. Denote the Pauli set of
qubit $(i,$ $j)$ as $\mathbb{P}_{(i,\text{ }j)}=\{X_{(i,\text{ }j)}$,
$Y_{(i,\text{ }j)}$, $Z_{(i,\text{ }j)}$, $I_{(i,\text{ }j)}\}$, where
$X_{(i,\text{ }j)}=\sigma_{x},$ $Y_{(i,\text{ }j)}=\sigma_{y},$ $Z_{(i,\text{
}j)}=\sigma_{z}$, and $I_{(i,\text{ }j)}=I$. Let $h_{(i)}$ $\in\{\overline
{X}_{(i)}^{\text{ }},$ $g_{(i)}^{\prime}\}$ and $h_{(i)}^{\prime}$
$\in\{\overline{X}_{(i)}^{\text{ }},$ $\overline{Z}_{(i)}^{\text{ }}\}$. Note
that
\begin{equation}
\left[  g_{(i)}\text{, }h_{(i)}\right]  =\left[  g_{(i)}\text{, }%
h_{(i)}^{\prime}\right]  =0, \label{commu}%
\end{equation}
and since $g_{(i)}$, $h_{(i)}$, and $h_{(i)}^{\prime}$ are $n_{i}$-fold tensor
products of the Pauli operators, we have
\begin{equation}
g_{(i)}=\prod_{j=1}^{n_{i}}\widehat{s}_{(i,\text{ }j)},h_{(i)}=\prod
_{j=1}^{n_{i}}\widehat{t}_{(i,\text{ }j)},h_{(i)}^{\prime}=\prod_{j=1}^{n_{i}%
}\widehat{t}_{(i,\text{ }j)}^{\prime},
\end{equation}
where $\widehat{s}_{(i,\text{ }j)}$, $\widehat{t}_{(i,\text{ }j)}$, and
$\widehat{t}_{(i,\text{ }j)}^{\prime}\in\mathbb{P}_{(i,\text{ }j)}$ $\forall
i$, $j.$

Before proceeding further, some notations are introduced as follows: Denote
the qubit index sets as $\mathbb{D}_{i}=\{j|\{\widehat{s}_{(i,\text{ }%
j)},\widehat{t}_{(i,\text{ }j)}\}=0\}$ and $\mathbb{H}_{i}=\{j|[\widehat
{s}_{(i,\text{ }j)},\widehat{t}_{(i,\text{ }j)}]=0\}$, where $\mathbb{D}%
_{i}\cap\mathbb{H}_{i}=\emptyset$, and $\left\vert \mathbb{D}_{i}\right\vert
+\left\vert \mathbb{H}_{i}\right\vert =n_{i}$. The qubits belonging to the
sets $\mathbb{D}_{1}$, $\cdots$, $\mathbb{D}_{N}$ play substantial roles in
the proposed Bell inequalities of the quantum networks. Note that qubit $(i$,
$j)$ is called idle if $(i$, $j)\in$ $\mathbb{H}_{i}$ and $(i$, $j)\rightarrow
\mathcal{R}^{(k)}$ for some $k$, and the nonidentity operator of $\widehat
{s}_{(i,\text{ }j)}\ $or $\widehat{t}_{(i,\text{ }j)}$ is denoted by
$\widehat{o}_{(i,\text{ }j)}$, which will be repeatedly measured on qubit
$(i$, $j)$ in the Bell test. Let the indicator $\delta_{(i,\text{ }j)}^{D}=1$
if qubit $(i$, $j)\in$ $\mathbb{D}_{i}$, and $0$ if $(i$, $j)\in$
$\mathbb{H}_{i}$. According to (\ref{commu}), we have $\sum_{j=1}^{n_{i}%
}\delta_{(i,\text{ }j)}^{D}\operatorname{mod}2=0,$ and hence,
\begin{equation}
\sum_{i=1}^{N}\sum_{j=1}^{n_{i}}\delta_{(i,\text{ }j)}^{D}\operatorname{mod}%
2=0. \label{t1}%
\end{equation}

Here, we focus on quantum networks fulfilling the following conditions:
Regarding the qubits held by $\mathcal{S}^{(k)}$,%

\begin{equation}
\sum\nolimits_{i=e_{k-1}+1}^{e_{k}}\sum\nolimits_{j=1,(i,\text{ }%
j)\in\mathbb{S}^{(k)}}^{n_{i}}\delta_{(i,\text{ }j)}^{D}\operatorname{mod}%
2=1\text{ }\forall k=1,2,\cdots,K\text{,} \label{t2}%
\end{equation}
and regarding the qubits held by $\mathcal{R}^{(k)}$,%

\begin{equation}
\sum\nolimits_{i,\text{ }j,(i,\text{ }j)\in\mathbb{R}^{(k)}}\delta_{(i,\text{
}j)}^{D}\operatorname{mod}2=1\text{ }\forall j=1,\cdots,M. \label{t3}%
\end{equation}
Combining the constraints (\ref{t1}), (\ref{t2}), and (\ref{t3}), the value
$K+M$ must be even. According to the qubit distribution, it is very flexible
to choose suitable $g_{(i)}$, $h_{(i)}$, and $h_{(i)}^{\prime}$ to implement
local observables.

\subsection{Violation of the Bell inequalities in a quantum network}

To implement the local observables on the source side, we assign%
\begin{equation}
\widehat{S}^{(k)}=\prod\nolimits_{i,\text{ }j,\text{ }(i,\text{ }%
j)\in\mathbb{S}^{(k)}}\widehat{s}_{(i,\text{ }j)}\rightarrow\frac{1}{2\cos\mu
}(A_{0}^{(k)}+A_{1}^{(k)}) \label{SS}%
\end{equation}
and
\begin{equation}
\widehat{T}^{(k)}=\prod\nolimits_{i^{\prime},\text{ }j^{\prime},\text{
}(i^{\prime},\text{ }j^{\prime})\in\mathbb{S}^{(k)}}\widehat{t}_{(i^{\prime
},\text{ }j^{\prime})}\rightarrow\frac{1}{2\sin\mu}(A_{0}^{(k)}-A_{1}^{(k)}).
\label{TT}%
\end{equation}
Since $A_{x_{k}}^{(k)}=A_{x_{k}}^{(k)\dag}$and $\left(  A_{xi}^{(i)}\right)
^{2}=A_{x_{i}}^{(i)}A_{xi}^{(i)\dag}=I$, $A_{x_{k}}^{(k)}$ is a unitary
Hermitian with an eigenvalue of either $1$ or $-1$. On the receiving side, the
local observables $B_{y_{l}}^{(l)}$ for observer $\mathcal{R}^{(k)}$ are%

\begin{equation}
B_{y_{l}}^{(l)}=(1-y_{l})\prod\nolimits_{i,\text{ }j,\text{ }(i,\text{ }%
j)\in\mathbb{R}^{(l)}}\widehat{s}_{(i,\text{ }j)}+y_{l}\prod
\nolimits_{i^{\prime},\text{ }j^{\prime},\text{ }(i^{\prime},\text{ }%
j^{\prime})\in\mathbb{R}^{(l)}}\widehat{t}_{(i^{\prime},\text{ }j^{\prime})},
\end{equation}
where $\left\{  B_{0}^{(k)}\text{, }B_{1}^{(k)}\right\}  =0$ according to
(\ref{t3}). In the measuring phase, $\mathcal{S}^{(k)}$ ($\mathcal{R}^{(k)}$ )
randomly measures either the observable $A_{0}^{(k)}$ or $A_{1}^{(k)}$
($B_{0}^{(k)}$ or $B_{1}^{(k)}$) with an outcome of either $1$ or $-1$. In
practice, if the qubit $(i^{\prime\prime},$ $j^{\prime\prime})\rightarrow
\mathcal{R}^{(k)}$ is idle, $\mathcal{R}^{(k)}$ can always measure the
observable $\widehat{o}_{(i^{\prime\prime},j^{\prime\prime})}$ in each round
and discard the outcome in postprocessing if $\widehat{s}_{(i^{\prime\prime
},\text{ }j^{\prime\prime})}=I$ or $\widehat{t}_{(i^{\prime\prime},\text{
}j^{\prime\prime})}=I$.

To evaluate the correlation strength, let $\left\vert \overline{\Psi
}\right\rangle $ $=\prod\nolimits_{i\text{ }}^{N}\otimes\left\vert \psi
_{(i)}\right\rangle ,$ and we have%

\begin{align}
&  \left\vert \mathbf{I}_{K,M}\right\vert _{Q}\nonumber\\
&  =\frac{1}{2^{K}}\left\vert \sum_{x_{1},\cdots x_{M}=0}^{1}\left\langle
A_{x_{1}}^{(1)}\cdots A_{x_{K}}^{(K)}\prod_{j=1}^{M}B_{0}^{(j)}\right\rangle
\right\vert \nonumber\\
&  =\frac{1}{2^{K}}\left\vert \left\langle \prod_{i=1}^{K}(A_{0}^{(i)}%
+A_{1}^{(i)})\prod_{j=1}^{M}B_{0}^{(j)}\right\rangle \right\vert \nonumber\\
&  =\left\vert \left\langle \prod_{k=1}^{K}\cos\theta_{k}\prod
\nolimits_{i,\text{ }j,\text{ }(i,\text{ }j)\in\mathbb{S}^{(k)}}\widehat
{s}_{(i,\text{ }j)}\prod_{l=1}^{M}\prod\nolimits_{i^{\prime},\text{ }%
j^{\prime},\text{ }(i^{\prime},\text{ }j^{\prime})\in\mathbb{R}^{(l)}}%
\widehat{s}_{(i^{\prime},\text{ }j^{\prime})}\right\rangle \right\vert
\nonumber\\
&  =\left\vert \prod_{k=1}^{K}\cos\theta_{k}\left\langle \prod_{i=1}%
^{N}g_{(i)}\right\rangle \right\vert \nonumber\\
&  =\left\vert \prod_{k=1}^{K}\cos\theta_{k}\right\vert ,
\end{align}
where $\left\langle \cdot\right\rangle =\left\langle \overline{\Psi}\left\vert
\cdot\right\vert \overline{\Psi}\right\rangle $ and hence $\left\langle
\prod_{i=1}^{N}g_{(i)}\right\rangle =1$. \bigskip\ In addition,%
\begin{align}
&  \left\vert \mathbf{J}_{K,M}\right\vert _{Q}\nonumber\\
&  =\frac{1}{2^{K}}\left\vert \sum_{x_{1},\cdots x_{M}=0}^{1}(-1)^{\sum x_{j}%
}\left\langle A_{x_{1}}^{(1)}\cdots A_{x_{n}}^{(n)}\prod_{j=1}^{M}B_{1}%
^{(j)}\right\rangle \right\vert \nonumber\\
&  =\frac{1}{2^{K}}\left\vert \left\langle \prod_{i=1}^{K}(A_{0}^{(i)}%
-A_{1}^{(i)})\prod_{j=1}^{M}B_{1}^{(j)}\right\rangle \right\vert \nonumber\\
&  =\left\vert \left\langle \prod_{k=1}^{K}\sin\theta_{k}\prod
\nolimits_{i,\text{ }j,\text{ }(i,\text{ }j)\in\mathbb{S}^{(k)}}\widehat
{t}_{(i,\text{ }j)}\prod_{l=1}^{M}\prod\nolimits_{i^{\prime},\text{ }%
j^{\prime},\text{ }(i^{\prime},\text{ }j^{\prime})\in\mathbb{R}^{(l)}}%
\widehat{t}_{(i^{\prime},\text{ }j^{\prime})}\right\rangle \right\vert
\nonumber\\
&  =\left\vert \prod_{j=1}^{K}\sin\theta_{j}\prod_{i=1}^{N}\left\langle
h_{(i)}\right\rangle \right\vert .
\end{align}
A useful lemma is introduced as follows:

\bigskip\textit{Lemma 2 }For any $\theta_{1},\theta_{2},\cdots,\theta_{K}%
\in\lbrack0,\frac{\pi}{2}]$,
\begin{equation}
\left\vert \prod\nolimits_{j=1}^{K}\sin\theta_{j}\right\vert ^{\frac{1}{K}%
}\leq\sin\overline{\theta}\text{ and }\left\vert \prod\nolimits_{j=1}^{K}%
\cos\theta_{j}\right\vert ^{\frac{1}{K}}\leq\cos\overline{\theta},
\end{equation}
where $\overline{\theta}=\frac{1}{K}\sum_{j=1}^{K}\theta_{j}$. The proof can
be found in \cite{luo1}.

Let $c_{i}=\left\langle h_{(i)}\right\rangle \leq1$ and $C=\left\vert
\prod_{i=1}^{N}c_{i}\right\vert ^{\frac{1}{K}}$. We have%

\begin{align}
&  \left\vert \mathbf{I}_{K,M}\right\vert _{Q}^{\frac{1}{K}}+\left\vert
\mathbf{J}_{K,M}\right\vert _{Q}^{\frac{1}{K}}\nonumber\\
&  =\left\vert \prod\nolimits_{k=1}^{K}\cos\theta_{j}\right\vert ^{\frac{1}%
{K}}+\left\vert \prod\nolimits_{k=1}^{K}\sin\theta_{j}\right\vert ^{\frac
{1}{K}}\left\vert C\right\vert \nonumber\\
&  \leq\left\vert \cos\overline{\theta}\right\vert +\left\vert \sin
\overline{\theta}\right\vert \left\vert C\right\vert \nonumber\\
&  \leq\sqrt{1+C^{2}}, \label{Max}%
\end{align}
where the first inequality in (\ref{Max}) follows from Lemma 2 and the second
inequality follows from the Cauchy-Schwarz inequality. To reach the maximal
violation, by setting $\theta_{1}=\ldots=\theta_{K}=\overline{\theta}$, and
$\tan\overline{\theta}=C$, we obtain
\begin{equation}
\max(\left\vert \mathbf{I}_{K,M}\right\vert _{Q}^{\frac{1}{K}}+\left\vert
\mathbf{J}_{K,M}\right\vert _{Q}^{\frac{1}{K}})=\sqrt{1+C^{2}}>1.
\label{Violation}%
\end{equation}

Some discussion is in order.\ The state $\left\vert \psi_{(i)}\right\rangle $
can contribute to the nonlocality of the quantum network through different
facets of its nonlocality. Loosely speaking, there are $C_{2}^{2^{n_{i}-k_{i}%
}}$ ways of choosing suitable $\ g_{(i)}$ and $g_{(i)}^{\prime}$ and
$k_{i}(2^{2(n_{i}-k_{i})})$ ways of choosing suitable $\ g_{(i)}$ and
$\overline{X}_{(i)}^{\text{ }}$, which reflect the flexibility of testing Bell
nonlocality in the quantum networks. The maximum value of $C=1$ ($\overline
{\theta}=\frac{\pi}{4}$), and hence $\max(\left\vert I_{K,M}\right\vert
_{Q}^{\frac{1}{K}}+\left\vert J_{K,M}\right\vert _{Q}^{\frac{1}{K}})=\sqrt{2}%
$, can be reached if $c_{i}=1$ $\forall$ $i$. That is,\ $h_{(i)}$\ stabilizes
$\left\vert \psi_{(i)}\right\rangle $ $\forall$ $i$. As shown in (\ref{SS})
and (\ref{TT}), one can benefit from the stabilizing operators in designing
local observables to achieve the maximal violations of Bell inequalities in
$K$-locality quantum networks. In detail, $c_{i}=1$ if $h_{(i)}=$
$\overline{X}_{(i)}^{\text{ }}$, $\left\vert \left\langle \varphi_{i}%
^{0}|\varphi_{i}^{1}\right\rangle \right\vert =1$, and $\phi_{i}=\frac{\pi}%
{4}$. The Bell nonlocality here is due to the superposition of $\left\vert
\overline{0}_{i}\right\rangle $ and $\left\vert \overline{1}_{i}\right\rangle
$; if $h_{(i)}=$ $g_{(i)}^{\prime}$, $c_{i}$ is certain to be $1$. The Bell
nonlocality in a network is due to that of either the stabilizer state or the
logical states of the stabilizer code [[$n_{i}$, $k_{i}$, $d_{i}$]]. Note that
such nonlocality involving the stabilizing operators $g_{(i)}$ and
$g_{(i)}^{\prime}$ can be also obtained even by using specific states
stabilized by the same stabilizing operators. For example, given source $i$
emitting a four-qubit state with stabilizing operators $g_{(i)}=$ $\sigma
_{z}^{\otimes4}$ and $g_{(i)}^{\prime}=$ $\sigma_{x}^{\otimes4}$, which both
can also stabilize the 4-qubit Smolin state $\rho_{\text{Smolin}}=\frac{1}%
{4}\sum\nolimits_{ij=0,1}\left\vert \Psi_{ij}\right\rangle \left\langle
\Psi_{ij}\right\vert $, where $\left\vert \Psi_{ij}\right\rangle =\sigma
_{z}^{i}\otimes\sigma_{x}^{j}[\frac{1}{\sqrt{2}}(\left\vert 00\right\rangle
+\left\vert 11\right\rangle )]$. Eventually, Bell nonlocality of at most
$\left\vert \mathbb{D}_{i}\right\vert $-qubit entanglement in $\left\vert
\psi_{(i)}\right\rangle $ is involved.

As an illustration, let $K=N=2$ and $M=1$, and let sources 1 and 2 each emit
the codeword states of $[[5$, $1$, $3]]$ SQECC with four stabilizer generators
(the subscript qubit index $(i,$ $j)$ is shortened as $j$)
\begin{align}
\mathfrak{g}_{1}  &  =X_{1}Z_{2}Z_{3}X_{4}I_{5},\mathfrak{g}_{2}=I_{1}%
X_{2}Z_{3}Z_{4}X_{5},\nonumber\\
\mathfrak{g}_{3}  &  =X_{1}I_{2}X_{3}Z_{4}Z_{5},\mathfrak{g}_{4}=Z_{1}%
X_{2}I_{3}X_{4}Z_{5}. \label{gg}%
\end{align}
The well-known logical bit-flip and phase-flip operators are $\overline
{X}^{\prime\text{ }}=\prod\nolimits_{i=1}^{5}X_{i}$ and $\overline{Z}%
^{\prime\text{ }}=\prod\nolimits_{i=1}^{5}Z_{i}$. Observers $\mathcal{S}%
^{(1)}$ and $\mathcal{S}^{(2)}$ hold qubits $(1,$ $1)$ and $(2,$ $1)$,
respectively, while observer $\mathcal{R}^{(k)}$ holds the other 8 qubits, as
shown in Fig. (2). Note that, in the following examples, the bilocal
inequality is
\begin{equation}
\sqrt{\mathbf{I}_{2,1}}+\sqrt{\mathbf{J}_{2,1}}\leq1, \label{bibi}%
\end{equation}
which is exactly the bilocal inequality for binary inputs and outputs in
\cite{bilocal1}.

\textit{Example (a)}:\textit{ }Let $\left\vert \psi_{(1)}\right\rangle
=\left\vert \psi_{(2)}\right\rangle =\cos\phi\left\vert \overline
{0}\right\rangle +\sin\phi\left\vert \overline{1}\right\rangle $. Here, we
choose the useful operators $g_{(i)}=\mathfrak{g}_{1}\mathfrak{g}%
_{2}\mathfrak{g}_{3}\mathfrak{g}_{4}=Z_{1}Z_{2}X_{3}I_{4}X_{5}$ and
$h_{(i)}=\overline{X}^{\text{ }}=\overline{X}^{\prime}$. In this case,
$\left\vert \mathbb{D}_{i}\right\vert =2$, and $(i,$ $3)$, $(i,$ $4)$, $(i,$
$5)$ are idle qubits. $A_{x_{i}}^{(i)}=\frac{1}{\sqrt{2}}(Z_{(i,1)}%
+(-1)^{x_{i}}X_{(i,1)})$, $B_{0}^{(1)}=\prod\nolimits_{i=1}^{2}Z_{(i,2)}%
X_{(i,3)}X_{(i,5)}$, and $B_{1}^{(1)}=\prod\nolimits_{i=1}^{2}X_{(i,2)}%
X_{(i,3)}X_{(i,4)}X_{(i,5)}$. Note that $\mathbf{I}_{2,1}$ and $\mathbf{J}%
_{2,1}$ here are formally equivalent to $I^{22}$ and $J^{22}$ in
[\cite{bilocal1}]. \ Denote the 5-qubit state $\left\vert \varphi\right\rangle
=(\cos\phi\left\vert 0\right\rangle _{1}\left\vert 0\right\rangle _{2}%
+\sin\phi\left\vert 1\right\rangle _{1}\left\vert 1\right\rangle
_{2})\left\vert +\right\rangle _{3}\left\vert +\right\rangle _{4}\left\vert
+\right\rangle _{5}$. It is easy to verify that $g_{(i)}\left\vert
\varphi\right\rangle =\left\vert \varphi\right\rangle $ and $\left\langle
\psi_{(i)}\left\vert h_{(i)}\right\vert \psi_{(i)}\right\rangle =\left\langle
\varphi\left\vert h_{(i)}\right\vert \varphi\right\rangle $. As a result, the
same violation can be obtained using either $\left\vert \psi_{(1)}%
\right\rangle \left\vert \psi_{(2)}\right\rangle $ or $\left\vert
\varphi\right\rangle ^{\otimes2}$.

\textit{Example (b)}:\textit{ }Let $\left\vert \psi_{(1)}\right\rangle
=\left\vert \overline{0}\right\rangle $, $\left\vert \psi_{(2)}\right\rangle
=\left\vert \overline{1}\right\rangle $. Here, we choose the useful operators
$g_{(i)}=\mathfrak{g}_{1}\mathfrak{g}_{2}\mathfrak{g}_{3}\mathfrak{g}%
_{4}=Z_{1}Z_{2}X_{3}I_{4}X_{5}$ and $h_{(i)}=\mathfrak{g}_{1}$ for any
$i\in\{1,2\}$. In this case, the local observables $A_{x_{i}}^{(i)}$ and
$B_{0}^{(1)}$ are the same as those in example (a), while $B_{1}^{(1)}%
=\prod\nolimits_{i=1}^{2}Z_{(i,2)}Z_{(i,3)}X_{(i,4)}$. Here, $\left\vert
\mathbb{D}_{i}\right\vert =2$ and $(i,$ $2)$, $(i,$ $4)$, $(i,$ $5)$ are idle
qubits. However, note that $g_{(i)}\left\vert \varphi^{\prime}\right\rangle
=h_{(i)}\left\vert \varphi^{\prime}\right\rangle =\left\vert \varphi^{\prime
}\right\rangle $, where $\left\vert \varphi^{\prime}\right\rangle =\frac
{1}{\sqrt{2}}(\left\vert 0\right\rangle _{1}\left\vert +\right\rangle
_{3}+\left\vert 1\right\rangle _{1}\left\vert -\right\rangle _{3})\left\vert
0\right\rangle _{2}\left\vert +\right\rangle _{4}\left\vert +\right\rangle
_{5}$. The maximal violation can be obtained using either $\left\vert
\psi_{(1)}\right\rangle \left\vert \psi_{(2)}\right\rangle $ or $\left\vert
\varphi^{\prime}\right\rangle ^{\otimes2}$.

Consequently, although the two 5-qubit two logical states $\left\vert
\overline{0}\right\rangle $and $\left\vert \overline{1}\right\rangle $ and the
codeword states are genuinely entangled \cite{Add, bdd}, one can replace the
genuinely entangled state with a state involving two-qubit entanglement,
either $\left\vert \varphi\right\rangle $ or $\left\vert \varphi^{\prime
}\right\rangle $, to reach the same correlation strength. It is not known
whether the Bell nonlocality of genuine entanglement can be revealed in a
$K$-locality quantum network. Specifically, it is not known whether the
genuine entanglement of $\left\vert \psi_{(i)}\right\rangle $ can be deduced
from violations of variant Bell inequalities of a quantum network involving
different stabilizing operators.

\section{Bell inequalities tailored for nonmaximal entangled states in a
quantum network}

If $c_{i}<1$ for at least one $i$, $\left\vert \overline{\Psi}\right\rangle $
cannot achieve the maximal violation of the nonlinear Bell inequality
(\ref{NonBell}). To explore the Bell inequalities maximally violated by
$\left\vert \overline{\Psi}\right\rangle $, recall that in the two-qubit case
($N=K=M=1$, $n_{1}=2$), the tilted CHSH operator $\mathbf{B}_{\beta
\text{-}CHSH}=\beta B_{0}+$ $\mathbf{B}_{CHSH}$ is exploited using the logical
phase-flip operator $B_{0}$ with appropriate parameter $\beta$. Although it is
unlikely that $\prod_{m=1}^{M}B_{0}^{(m)\text{ }}=$ $\prod_{i=1}^{N}%
\overline{Z}_{(i)}^{\text{ }}$ in quantum networks, it will be shown that the
logical phase-flip operators are still useful in finding the tilted Bell
inequalities. Denote two index sets $\mathfrak{C}=\{i|c_{i}=1\}$ and
$\mathfrak{c}=\{i^{\prime}|c_{i^{\prime}}<1\}$, where $\mathfrak{C\cap
c=}\emptyset$ and $\left\vert \mathfrak{C}\right\vert +\left\vert
\mathfrak{c}\right\vert =N$. Without loss of generality, let $i\in
\mathfrak{C}$ for $1\leq i\leq\left\vert \mathfrak{C}\right\vert $ and
$i\in\mathfrak{c}$ for $\left\vert \mathfrak{C}\right\vert +1\leq
i\leq\left\vert \mathfrak{C}\right\vert +\left\vert \mathfrak{c}\right\vert
=N$. Denote $\left\vert \overline{\Psi}_{\mathfrak{C}}\right\rangle
=\prod\nolimits_{i=1\text{ }}^{\left\vert \mathfrak{C}\right\vert }\left\vert
\psi_{(i)}\right\rangle $ and $\left\vert \overline{\Psi}_{\mathfrak{c}%
}\right\rangle =\prod\nolimits_{i=\left\vert \mathfrak{C}\right\vert +1\text{
}}^{N}\left\vert \psi_{(i)}\right\rangle $; hence, we have $\left\vert
\overline{\Psi}\right\rangle =\left\vert \overline{\Psi}_{\mathfrak{C}%
}\right\rangle \otimes$ $\left\vert \overline{\Psi}_{\mathfrak{c}%
}\right\rangle $. Given $i^{\prime}\in\mathfrak{c}$, set the suitable operator
$h_{(i^{\prime})}^{\prime}$ $=$ $\overline{Z}_{(i^{\prime})}^{\text{ }}%
=\prod_{j^{\prime}=1}^{n_{i^{\prime}}}\widehat{t^{\prime}}_{(i^{\prime},\text{
}j^{\prime})}$ that fulfills the following conditions: (i) if $(i^{\prime},$
$j^{\prime})\rightarrow\mathcal{S}^{(k)},$ then $\widehat{t^{\prime}%
}_{(i^{\prime},\text{ }j^{\prime})}=I;$ and (ii) if $(i^{\prime},$ $j^{\prime
})\rightarrow\mathcal{R}^{(k)}$, then either qubit $(i^{\prime},$ $j^{\prime
})$ is idle or $\widehat{t^{\prime}}_{(i^{\prime},\text{ }j^{\prime}%
)}=\widehat{s}_{(i^{\prime},\text{ }j^{\prime})}$ if qubit $(i^{\prime},$
$j^{\prime})$ is not idle. The logical phase-flip operator can be revised as
\begin{equation}
\overline{Z}_{(i^{\prime})}^{\text{ }}=\underset{j^{\prime},\text{ }%
(i^{\prime},\text{ }j^{\prime})\text{: on the source side }}{\prod}\widehat
{s}_{(i^{\prime},\text{ }j^{\prime})}\prod_{j^{\prime\prime},\text{
}(i^{\prime},\text{ }j^{\prime\prime}):\text{idle}}\widehat{o^{\prime}%
}_{(i^{\prime},\text{ }j^{\prime\prime})}, \label{Zop}%
\end{equation}
where $\widehat{o^{\prime}}_{(i^{\prime},\text{ }j^{\prime\prime})}%
\in\{\widehat{o}_{(i^{\prime},\text{ }j^{\prime\prime})}$, $I\}$. Instead of
measuring $B_{0}^{(k)\text{ }}$directly, the agent $\mathcal{R}^{(k)}$
measures
\begin{equation}
\overline{B}_{0}^{(k)\text{ }}=B_{0}^{(k)\text{ }}\underset{\text{ }%
}{\underset{\widehat{s}_{(i^{\prime},\text{ }j^{\prime})}=I}{\prod
_{(i^{\prime},\text{ }j^{\prime})\rightarrow\mathcal{R}^{(k)}}}}%
\widehat{t^{\prime}}_{(i^{\prime},j^{\prime})}. \label{BB0}%
\end{equation}
The outcome of $B_{0}^{(k)\text{ }}$can be obtained from that of $\overline
{B}_{0}^{(k)\text{ }}$by dropping any outcome of the local observables
$\widehat{t^{\prime}}_{(i^{\prime},j^{\prime})}$ in (\ref{BB0}); the outcome
of $\prod_{i^{\prime},i^{\prime}\in\mathfrak{c}}\overline{Z}_{(i^{\prime}%
)}^{\text{ }}$ can be obtained from that of $\prod_{k=1}^{K}\overline{B}%
_{0}^{(k)\text{ }}$by dropping any outcome of the qubit $(i^{\prime},$
$j^{\prime\prime})$ if (i) $i^{\prime}\in\mathfrak{C}$ or (ii) $i^{\prime}%
\in\mathfrak{c}$ and $\widehat{o^{\prime}}_{(i^{\prime},\text{ }%
j^{\prime\prime})}$ in (\ref{Zop}) is the unit matrix. In this case, we
propose the tilted Bell inequalities tailored for $\left\vert \overline{\Psi
}\right\rangle $%

\begin{equation}
G_{N,K}^{\beta}=\beta\left\vert P_{N,K}\right\vert ^{\frac{1}{K}}+\left\vert
I_{N,K}\right\vert ^{\frac{1}{K}}+\left\vert J_{N,K}\right\vert ^{\frac{1}{K}%
}\overset{c}{\leq}\beta+1,
\end{equation}
where \bigskip$P_{N,K}=\prod\nolimits_{i=1\text{ }}^{\left\vert \mathfrak{C}%
\right\vert }I^{\otimes n_{i}}\prod\nolimits_{i^{\prime}=\left\vert
\mathfrak{C}\right\vert +1\text{ }}^{N}\overline{Z}_{(i^{\prime})}^{\text{ }}%
$. However, it is very difficult to exploit sum-of-squares decomposition to
find the tailored Bell operators in quantum networks with an extremely large
Hilbert space \cite{non2, SOS1}. Our strategy is to simplify $G_{N,K}^{\beta}$
as tilted CHSH inequalities. In detail, according to Lemma 2, we have
\begin{equation}
G_{N,K}^{\beta}\leq\overline{G}_{N,K}(\beta,\overline{\phi},\overline{\theta
})=\beta(\cos2\overline{\phi})^{\frac{\left\vert \mathfrak{c}\right\vert }{K}%
}+\cos\overline{\theta}+\sin\overline{\theta}(\sin2\overline{\phi}%
)^{\frac{\left\vert \mathfrak{c}\right\vert }{K}},
\end{equation}
where equality holds when $\theta_{1}=\ldots=\theta_{N}=\overline{\theta}$.
Let the parameter $\beta$ satisfying $\frac{\partial}{\partial\overline
{\theta}}\overline{G}_{N,K}|_{(\overline{\theta}_{\max},\overline{\phi}%
)}=\frac{\partial}{\partial\overline{\phi}}\overline{G}_{N,K}|_{(\overline
{\theta}_{\max},\overline{\phi})}=0$ be $\beta_{\max}$. We have
\begin{equation}
\tan\overline{\theta}_{\max}=(\sin2\overline{\phi})^{\frac{\left\vert
\mathfrak{c}\right\vert }{K}}, \label{tang}%
\end{equation}
and
\begin{equation}
\beta_{\max}=\frac{(\tan2\overline{\phi})^{^{\frac{2\left\vert \mathfrak{c}%
\right\vert }{K}-2}}}{\sqrt{(1+\tan^{^{2}}2\overline{\phi})^{\frac{\left\vert
\mathfrak{c}\right\vert }{K}}+(\tan2\overline{\phi})^{\frac{2\left\vert
\mathfrak{c}\right\vert }{K}}}}. \label{betamax}%
\end{equation}
As a result, the state $\left\vert \overline{\Psi}_{\mathfrak{C}}\right\rangle
\otimes$ $\left\vert \overline{\Psi}_{\mathfrak{c}}\right\rangle $ can
maximally violate the tailored Bell inequality $G_{N,K}^{\beta_{\max}%
}=\overset{c}{\overline{G}_{N,K}(\beta_{\max},\overline{\phi},\overline
{\theta}_{\max})\leq}\beta_{\max}+1\ $by setting $\theta_{1}=\ldots=\theta
_{N}=\overline{\theta}_{\max}$. Note that $G_{N,K}^{\beta_{\max}}$ coincides
with the tilted CHSH inequality in the case that $\frac{\left\vert
\mathfrak{c}\right\vert }{K}=1$ \cite{non2}. \bigskip

As an example, we consider the following "star-like" quantum network as shown
in Fig. 2. Here, set $K=N$ as an odd integer and $M=1$. For any $i$, let
$\left\vert \psi_{(i)}\right\rangle =\cos\phi_{i}\left\vert \overline{0}%
_{i}\right\rangle +\sin\phi_{i}\left\vert \overline{1}_{i}\right\rangle $ be
the codeword of the $[[5$, $1$, $3]]$ QECC, where $i\in\mathfrak{C}$ if
$\phi_{i}=\frac{\pi}{4}$ and $i\in\mathfrak{c}$ if $\phi_{i}=\overline{\phi
}\neq\frac{\pi}{4}$. The useful operators are $g_{(i)}=\mathfrak{g}_{4}$,
$h_{(i)}=\overline{X}_{(i)}^{\prime\text{ }}$, and $h_{(i)}^{\prime}%
=\overline{Z}_{(i)}^{\text{ }}=\mathfrak{g}_{1}\mathfrak{g}_{3}\overline
{Z}_{(i)}^{\prime\text{ }}=-I_{(i,2)}X_{(i,3)}X_{(i,4)}I_{(i,5)}Z_{(i,1)}$. On
the source side, agent $\mathcal{S}^{(i)}$ possesses qubit $(i$, $2)$, and
using the cut-and-mix method, the local observables are set as $A_{x_{i}%
}^{(i)}=\cos\theta_{i}Z_{(i,\text{ }2)}+(-1)^{x_{i}}\sin\theta_{i}X_{(i,\text{
}2)}$ for any $i=1,..,N$. On the receiving side, the only agent $\mathcal{R}%
^{(1)}$ possesses qubits $(j$, $1)$, $(j$, $3),(j$, $4)$, and $(j$, $5)$,
$1\leq j\leq N$. Notably, $\widehat{t^{\prime}}_{(j,\text{ }2)}=I$,
$\widehat{t^{\prime}}_{(j,\text{ }1)}=$ $\widehat{s}_{(j,\text{ }1)}%
=\sigma_{z}$, and qubits $(j$, $3),(j$, $4)$, and $(j$, $5)$ are idle. Here
two local observables for $\mathcal{R}^{(1)}$ are $B_{0}^{(1)}=\prod
\nolimits_{j=1}^{N}X_{(j,3)}I_{(j,4)}X_{(j,5)}Z_{(j,1)}$ and $B_{1}%
^{(1)}=\prod\nolimits_{j=1}^{N}X_{(j,3)}X_{(j,4)}X_{(j,5)}X_{(j,1)}$. In this
case, one can set $\overline{B}_{0}^{(1)}=-B_{0}^{(1)}\prod\nolimits_{j=1}%
^{N}X_{(j,4)}$. In practice,$\ \mathcal{R}^{(1)}$ randomly measures
$\sigma_{x}$ or $\sigma_{z}$ on the qubits $(j$, $1)$, \ldots, $(N$, $1)$, and
always measures $\sigma_{x}$ on qubits $(j$, $3),(j$, $4)$, and $(j$, $5)$,
$1\leq j\leq N$. In this scenario, the numerical simulation shows that $\max
G_{N,K}^{\beta}=\beta_{\max}$.

\section{Conclusions}

In conclusion, we study the quantum networks with sources emitting different
stabilizer states. To characterize Bell nonlocality, knowledge of the emitted
entangled states is demonstrated to be quite useful. Regarding qubit
distributions in quantum networks, nonlinear Bell inequalities are proposed,
which can reveal variant facets of Bell nonlocality. On the other hand, by
fully exploiting the logical bit-flip and phase-flip operators, we derive
tilted nonlinear Bell inequalities tailored for the codewords of [[5, 1, 3]]
QECC with a specific qubit distribution. It is interesting to construct the
tilted nonlinear Bell inequalities maximally violated by $\left\vert
\overline{\Psi}\right\rangle $ comprising the generic codewords of QECCs in
quantum networks.

\section{Figure Captions}

Figure 1. The particle distribution of the state emitted from the source $i$
held by $\mathcal{S}^{(s)}$ if $e_{s-1}+1\leq i\leq e_{s}$. The particles
assigned on the source (receiving) side are inside (outside) the circle. The
state can be (a) the classical hidden state $\lambda_{i}$ or (b) the quantum
state $\left\vert \psi_{(i)}\right\rangle $. Here the particle ($i$, 1)
assigned on the source side is determined as the center of the star graph.
Qubit $(i,$ $j)$ in (b) is denoted by a black dot if $\{\widehat{s}_{(i,\text{
}j)},$ $\widehat{t}_{(i,\text{ }j)}\}=0$. A quantum network is depicted in
(c). The qubits in a rounded rectangle are locally accessible by an agent.
According to (\ref{t2}) and (\ref{t3}), there is an odd number of black dots
in each rounded rectangle.

Figure 2. A star-like quantum network with all states $\left\vert \psi
_{(1)}\right\rangle ,\ldots,\left\vert \psi_{(N)}\right\rangle $ being [[5, 1,
3]] codeword states.

\begin{thebibliography}{99}                                                                                               %


\bibitem {Bell}J. S. Bell, On the Einstein--Podolsky--Rosen paradox, Physics
\textbf{1} 19 (1964).

\bibitem {EPR}A. Einstein, B. Podolsky, and N. Rosen, Can Quantum-Mechanical
Description of Physical Reality Be Considered Complete?, Phys. Rev.
\textbf{47}, 777 (1935).

\bibitem {random}S. Pironio, A. Ac\'{\i}n, S. Massar, A. Boyer de la Giroday,
D. N. Matsukevich, P. Maunz, S. Olmschenk, D. Hayes, L. Luo, T. A.Manning, and
C. Monroe, Random numbers certified by Bell's theorem, Nature \textbf{464}%
,1021 (2010).

\bibitem {DIC}A. Ac\'{\i}n, N. Brunner, N. Gisin, S. Massar, S. Pironio, and
V. Scarani, Device-independent security of quantum cryptography against
collective attacks, Phys. Rev. Lett. \textbf{98}, 230501 (2007).

\bibitem {complexity}H. Buhrman, R. Cleve, S. Massar, and R. de Wolf,
Nonlocality and communication complexity, Rev. Mod. Phys. \textbf{82}, 665(2010).

\bibitem {gf1}M. Hein, J. Eisert, and H. J. Briegel, Multiparty entanglement
in graph states, Phys. Rev. A \textbf{69}, 062311 (2004).

\bibitem {gf2}L.-Y. Hsu, Bell-type inequalities embedded in the subgraph of
graph states, Phys. Rev. A \textbf{73}, 042308 (2006).

\bibitem {gf3}A. Cabello, O. G\"{u}hne, and D. Rodr\'{\i}guez, Mermin
inequalities for perfect correlations, Phys. Rev. A \textbf{77}, 062106 (2008).

\bibitem {gs2}V. Scarani, A. Ac\'{\i}n, E. Schenck, and M. Aspelmeyer,
Nonlocality of cluster states of qubits, Phys. Rev. A \textbf{71}, 042325 (2005).

\bibitem {gs3}M. Plesch and V. Bu\v{z}ek, Entangled graphs: Bipartite
entanglement in multiqubit systems, Phys. Rev. A \textbf{67}, 012322 (2003).

\bibitem {gs4}O. G\"{u}hne, G. T\'{o}th, P. Hyllus, and H. J. Briegel, Bell
Inequalities for graph States, Phys. Rev. Lett. \textbf{95}, 120405 (2005).

\bibitem {gs5}O. G\"{u}hne, A. Cabello, Generalized Ardehali-Bell inequalities
for graph states, Phys. Rev. A \textbf{77}, 032108 (2008).

\bibitem {mix1}G. Wang and M. Ying, Multipartite unlockable bound entanglement
in the stabilizer formalism, Phys. Rev. A \textbf{75}, 052332 (2007).

\bibitem {mix2}L.-Y. Hsu, Mixed entangled states with two or more common
stabilizers, Phys. Rev. A \textbf{76}, 022322 (2007).

\bibitem {non1}A. Ac\'{\i}n, S. Massar, and S. Pironio, Randomness versus
Nonlocality and Entanglement, Phys. Rev. Lett. \textbf{108}, 100402 (2012).

\bibitem {non2}C. Bamps and S. Pironio, Sum-of-squares decompositions for a
family of Clauser-Horne-Shimony-Holt-like inequalities and their application
to self-testing, Phys. Rev. A \textbf{91}, 052111(2015).

\bibitem {star}A. Tavakoli, P. Skrzypczyk, D. Cavalcanti, and A. Ac\'{\i}n,
Nonlocal correlations in the star-network configuration, Phys. Rev. A
\textbf{90}, 062109 (2014).

\bibitem {noncyclic}A. Tavakoli, Bell-type inequalities for arbitrary
noncyclic networks, Phys. Rev. A \textbf{93}, 030101(R) (2016).

\bibitem {Poly}R. Chaves, Polynomial Bell Inequalities, Phys. Rev. Lett.
\textbf{116}, 010402 (2016).

\bibitem {luo1}M.-X. Luo, Computationally effcient nonlinear Bell inequalities
for quantum networks, Phys. Rev. Lett. \textbf{120}, 140402 (2018).

\bibitem {bilocal1}C. Branciard, D. Rosset, N. Gisin, and S. Pironio, Bilocal
versus nonbilocal correlations in entanglement-swapping experiments, Phys.
Rev. A \textbf{85}, 032119 (2012).

\bibitem {bilocal2}C. Branciard, N. Brunner, H. Buhrman, R. Cleve, N. Gisin,
S. Portmann, D. Rosset, and M. Szegedy, Classical Simulation of Entanglement
Swapping with Bounded Communication, Phys. Rev. Lett. \textbf{109}, 100401 (2012).

\bibitem {luo2}M.-X. Luo, Nonlocality of all quantum networks, Phys. Rev. A
\textbf{98}, 042317 (2018).

\bibitem {Poly1}F. Andreoli1, G. Carvacho, L. Santodonato, R. Chaves and F.
Sciarrino, Maximal qubit violation of $n$-locality inequalities in a
star-shapedquantum network, New J. Phys. \textbf{19} 113020 (2017).

\bibitem {QSS}R. Cleve, D. Gottesman, and H.-K. Lo, How to share a quantum
secret, Phys. Rev. Lett. \textbf{83} 648 (1999).

\bibitem {Preskill}P. W. Shor and J. Preskill, Simple proof of security of the
BB84 quantum key distribution protocol, Phys. Rev. Lett. \textbf{85}, 441 (2000).

\bibitem {percolation}A. Ac\'{\i}n, I. Cirac, and M. Lewenstein, Entanglement
percolation in quantum networks, Nat. Phys. \textbf{3}, 256 (2007).

\bibitem {memory}K. Hammerer, A. S. S\H{o}ensen, and E. S. Polzik, Quantum
interface between light and atomic ensembles, Rev. Mod. Phys. \textbf{82},
1041 (2010).

\bibitem {repeater1}N. Sangouard, C. Simon, H. de Riedmatten, and N. Gisin,
Quantum repeaters based on atomic ensembles and linear optics, Rev. Mod. Phys.
\textbf{83}, 33 (2011).

\bibitem {QSS1}A. Mouzali, F. Merazka, and D. Markham, Quantum secret sharing
with error correction, Commun.Theor. Phys. \textbf{58 }661 (2012).

\bibitem {QSS2}Z.-R. Zhang, W.-T.Liu , and Cheng-Zu Li, Quantum secret sharing
based onquantum error-correcting codes, Chin. Phys. B \textbf{20} 050309 (2011).

\bibitem {Add}F. Baccari, R. Augusiak, I. \v{S}upi\'{c}, and A. Ac\'{\i}n,
Device-Independent Certification of Genuinely Entangled Subspaces Phys. Rev.
Lett. \textbf{125}, 260507 (2020).

\bibitem {bdd}F. Baccari, R. Augusiak, I. \v{S}upi\'{c}, J. Tura, and A.
Ac\'{\i}n, Scalable Bell Inequalities for Qubit Graph States and Robust
Self-Testing, Phys. Rev. Lett. \textbf{124}, 020402 (2020).

\bibitem {SOS1}D. Saha, R. Santos, and R. Augusiak, Sum-of-squares
decompositions for a family of noncontextuality inequalities and self-testing
of quantum devices, Quantum \textbf{4}, 302 (2020).
\end{thebibliography}
\end{document}